\documentclass[aps,prc,twocolumn,floatfix,superscriptaddress,nofootinbib]{revtex4-1}
\usepackage{graphicx}
\usepackage[pdftex,colorlinks,citecolor=blue,bookmarks]{hyperref}
\usepackage{bm}
\usepackage{amsmath}
\usepackage{longtable}
\usepackage{enumitem}
\usepackage{xcolor}

\begin{document}
\title{Optimization of the number of intrinsic states included in the discrete Generator Coordinate Method}
\author{Jaime Mart\'inez-Larraz} 
\email{jaime.martinez-larraz@uam.es}
\address{Departamento de F\'isica Te\'orica, Universidad Aut\'onoma de Madrid, E-28049 Madrid, Spain}
\author{Tom\'as R. Rodr\'iguez} 
\email{tr.rodriguez@ucm.es}
\address{Departamento de Estructura de la Materia, F\'isica T\'ermica y Electr\'onica, Universidad Complutense de Madrid, E-28040 Madrid, Spain}
\address{Departamento de F\'isica Te\'orica, Universidad Aut\'onoma de Madrid, E-28049 Madrid, Spain}
\address{Centro de Investigaci\'on Avanzada en F\'isica Fundamental-CIAFF-UAM, E-28049 Madrid, Spain}
\begin{abstract}
We present a mechanism to efficiently pre-select the number of intrinsic many-body states that are used to define the many-body wave functions within the discrete Generator Coordinate Method (GCM). This procedure, based on the proper definition of a natural basis of orthonormal states, does not require the evaluation of the non-diagonal Hamiltonian kernels to do the selection and helps to reduce the numerical instabilities. The performance of the method is analyzed in detail in the ground state and $0^{+}$ excited states of some selected nuclei computed with the Gogny energy density functional.  
\end{abstract}
\maketitle
\section{Introduction}
\label{Introduction}
The Generator Coordinate Method (GCM) provides a general framework to give variational solutions to the many-body problem~\cite{Hill53a,Griffin57a,RS80a}. It is based on the definition of the variational trial wave functions as the linear mixing of different intrinsic configurations defined along the so-called generating coordinates. This beyond-mean-field method can give ground and excitation energies, decay probabilities, and interpretations of the results in terms of collective and single-particle degrees of freedom. In nuclear physics, the most common (and involved) realizations of the GCM formalism nowadays is the mixing of symmetry-restored (particle-number, parity and angular momentum projected) intrinsic quasiparticle states obtained from self-consistent mean-field calculations. Such an implementation has been used in different contexts. Hence, with energy density functionals (EDF) it is often referred as multi-reference EDF (MR-EDF) or symmetry conserving configuration mixing (SCCM)~\cite{Bender03a,Lacroix09a,Niksic11a,Egido16a,Robledo18a}; with valence-space Hamiltonians is known as projected-GCM (PGCM)~\cite{Jiao17a,Bally19a,Sanchez_Fernandez21,Frosini22b,Frosini22c}, discretized non-orthogonal shell model (DNO-SM)~\cite{Dao22}, or Monte Carlo shell model (MCSM)~\cite{Shimizu21a}. All of these methods are very similar and they subtle differ by the definition of the intrinsic states, the generating coordinates used, the nuclear interaction, the selection of the states and/or a combination of the aforementioned. 

One of the properties of the GCM is that the variational approximation can be straightforwardly improved by adding more complexity to the trial wave functions, i.e., including more generating coordinates and/or quasiparticle configurations. Some methods incorporate a selection process to reduce the number of configurations that are actually used~\cite{Dao22,Shimizu21a,Romero21}. For example, in the MCSM the set of intrinsic states is chosen with a stochastic method and a new wave function is incorporated in the definition of the GCM state if the resulting diagonalization with such an augmented set produces an energy gain and a reduction of the dispersion of the energy. However, the computational burden can quickly become prohibitive because they involve the evaluation of costly matrix elements of the Hamiltonian. 

Another problem that the GCM presents is the practical solution of the Hill-Wheeler-Griffin (HWG) equation that is deduced from the variational method~\cite{Hill53a,Griffin57a,RS80a}. While the GCM was originally proposed as a continuous integral superposition, most modern applications use a discrete representation where the integral is replaced by a discrete sum of non-orthogonal states~\cite{Broeckhove1979}. Hence, the HWG equation is essentially a generalized eigenvalue problem that must be transformed first into a regular Schr\"odinger-like equation. This is normally performed by extracting an orthonormal and linear independent set of states, the so-called natural basis, from the original set of intrinsic states~\cite{Lowdin1956,Lathouwers1976a,Piza1977}. However, this basis presents numerical instabilities and the selection of the final results is usually done by analyzing the GCM energy as a function of the number of states included in the natural basis~\cite{Bonche90a,Robledo18a}. This selection and convergence of the results is sometimes delicate and, again, requires the evaluation of diagonal and non-diagonal Hamiltonian kernels. Additionally, the final number of states in the natural basis is smaller (or equal, at most) than the number of intrinsic states in the initial set due to exact and/or approximate linear dependencies in such a set~\cite{Lowdin1956,Lathouwers1976a,Lathouwers1976b}. Then, one could ask whether there is a limitation in the number of meaningful orthonormal states that can be obtained from an original set of wave functions that are built by exploring a given generating coordinate. 

In this work we study the convergence of the GCM energies with the number of intrinsic states defined along a generating coordinate within a pre-established interval. In particular, we perform calculations of three characteristic nuclei ($^{40}$Ca, $^{80}$Sr and $^{186}$Pb) with the Gogny D1S energy density functional~\cite{Berger84,Robledo18a} and the axial quadrupole deformation as the GCM degree of freedom. We analyze the properties of the different natural bases that can be built with more or less states included in such an original set and how these properties can help us to choose an optimal number of intrinsic states without computing costly Hamiltonian matrix elements at that stage of the calculation. The paper is organized as follows. First, we summarize the GCM method and propose a way to reduce the number of intrinsic configurations included in the calculations based on the orthonormality of the natural basis (Sec.~\ref{Theoretical_Framework}). Then, we analyze the results obtained with the Gogny EDF calculations in Sec.~\ref{Results} for $0^{+}$ ground and excited states. Finally, we summarize the main conclusions of the present work (Sec.~\ref{Summary}). 
\section{Theoretical framework}
\label{Theoretical_Framework}
We start with the definition of the discretized GCM ansatz~\cite{Broeckhove1979}. The nuclear many-body states are obtained as:
\begin{equation}
|\Psi^{\sigma}\rangle=\sum_{i=1}^{N_{\mathrm{int}}}f^{\sigma}_{q_{i}}|\Phi_{q_{i}}\rangle
\label{GCM_ansatz}
\end{equation}
where $\sigma=1,2,...$ is used to label the order of the states in the energy spectrum and $\lbrace |\Phi_{q_{i}}\rangle\rbrace_{i=1,..,N_{\mathrm{int}}}$ is a set of non-orthogonal -and possibly, linear dependent- intrinsic many-body states that have different values of the so-called generating coordinates, $\lbrace q_{i}\rbrace$, that are chosen on physical grounds. For example, in the PGCM, the set of intrinsic states is formed by particle number, parity and angular momentum projected Bogoliubov quasi-particle states, $|\Phi_{q_{i}}\rangle=P^{N} P^{Z} P^{\pi} P^{J}_{MK}  |\phi_{q_{i}}\rangle$. Here $P^{N(Z)}$, $P^{\pi}$, $P^{J}_{MK}$ are the projectors onto good number of neutrons (protons), parity and angular momentum and its projection onto the laboratory and body-fixed $z$-axis, respectively~\cite{Sheikh21a,Bally21b}. The states $|\phi_{q_{i}}\rangle$ are obtained by solving constrained Hartree-Fock-Bogoliubov (HFB) equations~\cite{RS80a} (or other versions like the variation after particle number projection, VAPNP~\cite{Anguiano01a}) with $\lbrace q_{i}\rbrace$ being multipole deformations, pairing content, cranking angular momentum, etcetera. For example, in this work we will use the axial quadrupole deformation, $\beta_{2}$, as the generating coordinate and is defined through the quadrupole moment, $Q_{20}=r^{2}Y_{20}(\theta,\varphi)$, as
\begin{equation}
\beta_{2} = \frac{4\pi Q_{20}}{3r_{0}^{2}A^{5/3}} 
\end{equation}
where $r_{0}=1.2$ fm, $A$ is the total mass number, and $r$ and $Y_{20}(\theta,\varphi)$ are the radial coordinate and the spherical harmonics of degree 2 and order $0$, respectively.

The application of the variational method to the coefficients of the linear combination in Eq.~\eqref{GCM_ansatz} gives the HWG equation:
\begin{equation}
\sum_{j=1}^{N_{\mathrm{int}}}\left(\mathcal{H}_{q_{i},q_{j}}-E^{\sigma}\mathcal{N}_{q_{i},q_{j}}\right)f^{\sigma}_{q_{j}}=0
\label{HWG_eq1}
\end{equation}
where the norm and Hamiltonian overlap matrices are defined as:
\begin{eqnarray}
\mathcal{N}_{q_{i},q_{j}}&=&\langle\Phi_{q_{i}}|\Phi_{q_{j}}\rangle\\
\mathcal{H}_{q_{i},q_{j}}&=&\langle\Phi_{q_{i}}|\hat{H}|\Phi_{q_{j}}\rangle
\end{eqnarray}
The solution of Eq.~\eqref{HWG_eq1} provides a variational approximation to the exact energies and wave functions of the nuclear Hamiltonian, $\hat{H}$. In fact, other observables and transition properties associated to a generic operator $\hat{O}$ can be evaluated through:
\begin{equation}
O^{\sigma_{1},\sigma_{2}}=\langle\Psi^{\sigma_{1}}|\hat{O}|\Psi^{\sigma_{2}}\rangle=\sum_{i,j=1}^{N_{\mathrm{int}}}f^{\sigma_{1}*}_{q_{i}}\langle\Phi_{q_{i}}|\hat{O}|\Phi_{q_{j}}\rangle f^{\sigma_{2}}_{q_{j}}
\end{equation}
We emphasize that the non-orthogonal condition of the intrinsic wave functions, and, more specifically, of the Bogoliubov quasiparticle states, is actually necessary to be able to use the Generalized Wick Theorem to compute non-diagonal matrix elements for the various operators~\cite{Balian69a}.
\subsection{Solution of the HWG equation}
The discrete HWG equation requires some processing to solve it. Hence, we have to define a proper orthonormal basis since the intrinsic states are not orthogonal and might be linear dependent. The usual way of finding such an orthonormal basis is through the canonical orthonormalization~\cite{Lathouwers1976a} where the eigenvalues $(\lambda_{k})$ and eigenvectors $(u_{\lambda_{k},q_{i}})$ of the norm overlap matrix are used to define the so-called natural basis, $\lbrace|\Lambda_{k}\rangle\rbrace_{k=1,..,N_{\mathrm{nat}}}$ with:
\begin{eqnarray}
\sum_{j=1}^{N_{\mathrm{int}}}\mathcal{N}_{q_{i},q_{j}}u_{\lambda_{k},q_{j}}&=&\lambda_{k}u_{\lambda_{k},q_{i}}\label{norm_over_mat}\\
|\Lambda_{k}\rangle&=&\sum_{i=1}^{N_{\mathrm{int}}} \frac{u_{\lambda_{k},q_{i}}}{\sqrt{\lambda_{k}}}|\Phi_{q_{i}}\rangle
\label{natural_basis}
\end{eqnarray}
The eigenvalues $\lbrace\lambda_{k}\geq0\rbrace$ for all $k$ since the norm overlap matrix is positive semi-definite. The most critical point of the stability of the method comes from the exact and/or approximate linear dependence (LD) of the intrinsic set of states. If there are $L_{\mathrm{exa}}$ intrinsic states such that
\begin{equation}
|\Phi_{q_{m}}\rangle=\sum_{i=1}^{N_{\mathrm{int}}-L_{\mathrm{exa}}}a_{i}|\Phi_{q_{i}}\rangle
\label{exact_ld}
\end{equation} 
then there will be $L_{\mathrm{exa}}$ eigenvalues with $\lambda_{m}$ exactly zero and the natural basis states coming from these eigenvalues/eigenvectors must not be taken into account ($m=N_{\mathrm{int}}-L_{\mathrm{exa}}+1,...,N_{\mathrm{int}}$). These kinds of states are found, e.g., in the $K$-mixing when angular momentum projection of signature symmetry conserving states is performed (see, e.g., Ref.~\cite{Bally21b}). 
However, there could be also $L_{\mathrm{app}}$ very small norm overlap matrix eigenvalues that do not correspond to intrinsic states fulfilling Eq.~\eqref{exact_ld}. On the contrary, increasingly small and different from zero eigenvalues may naturally appear in large positive definite matrices ($\lambda=0$ is an accumulation point for the eigenvalues~\cite{Lathouwers1976a}). This approximate LD may cause a numerically meaningless definition of several natural basis states as it was already pointed out by L. Lathouwers in the seventies of the past century~\cite{Lathouwers1976b}. Therefore, even if the number of elements of the natural basis should be equal to $N_{\mathrm{int}}-L_{\mathrm{exa}}$, we have to remove additionally $L_{\mathrm{app}}$ states that correspond to $\lambda_{k}<\varepsilon_{\lambda}$. It was also proven in Ref.~\cite{Lathouwers1976b} that the most faithful truncated natural basis to the original one corresponds to $\lbrace|\Lambda_{k}\rangle\rbrace_{k=1,..,N_{\mathrm{nat}}}$ with $N_{\mathrm{nat}}=N_{\mathrm{int}}-L_{\mathrm{exa}}-L_{\mathrm{app}}$ and using the $N_{\mathrm{nat}}$-largest eigenvalues of the norm overlap matrix to build and sort out such a basis, where $k=1$ corresponds to the largest eigenvalue, $k=2$ the second largest, and so on. The parameter $\varepsilon_{\lambda}$ must be a small number to be determined heuristically (see below). 

Instead of using $\varepsilon_{\lambda}$, we propose in the present work to use the deviation of the natural basis from its orthonormality as a more restricted and numerically stable method to determine the number of elements of the natural basis:
\begin{equation}
\langle\Lambda_{k}|\Lambda_{k'}\rangle-\delta_{kk'}<\varepsilon_{\mathrm{nat}}\,\,\,;\,\,\,\forall k,k'
\label{orth_cond}
\end{equation}  
That means that the states that do not fulfill Eq.~\eqref{orth_cond} are removed from the natural basis.

Once the natural basis is properly defined, the GCM many-body states can be now spanned in such a basis as:
\begin{equation}
|\Psi^{\sigma}\rangle=\sum_{k=1}^{N_{\mathrm{nat}}}g^{\sigma}_{k}|\Lambda_{k}\rangle
\end{equation}
and the HWG equation can be read as a normal eigenvalue problem:
\begin{equation}
\sum_{k'=1}^{N_{\mathrm{nat}}}\langle\Lambda_{k}|\hat{H}|\Lambda_{k'}\rangle g^{\sigma}_{k'}=E^{\sigma}g^{\sigma}_{k}
\label{HWG_2}
\end{equation}

The usual procedure to check the convergence of the GCM results is the study of the GCM energy, $E^{\sigma}(N_{\mathrm{nat}})$, obtained by solving Eq.~\eqref{HWG_2} with an increasing number of states in the natural basis~\cite{Bonche90a,Robledo18a}. If the contribution to the GCM wave function of the last natural basis states -with smaller norm overlap matrix eigenvalues- is increasingly less significant, then a nearly constant $E^{\sigma}(N_{\mathrm{nat}})$ within a range $N_{\mathrm{nat},_{\mathrm{min}}}\leq N_{\mathrm{nat}}\leq N_{\mathrm{nat},_{\mathrm{max}}}$. This is the so-called plateau condition. In fact, if we take $N_{\mathrm{nat}}> N_{\mathrm{nat},_{\mathrm{max}}}$, or, equivalently, $\varepsilon_{\lambda}$ very small, then the exact and/or approximate linear dependencies should become numerically evident in the form of a sudden jump in the energy. This is the best case scenario to identify the optimal $N_{\mathrm{nat}}$ and the converged GCM energy, $E^{\sigma}$. However, a more or less continuous decrease and/or small jumps before the LD breakdown is found in many practical calculations and sometimes is difficult to asses unambiguously the final value of the GCM energy and the wave function. 

Another interesting aspect also found in the study of the plateaux is the similar convergence of the energy with dense and sparse initial sets of intrinsic wave functions provided that these sets are defined along a given collective coordinate with the same boundaries, e.g., the axial quadrupole deformation $\beta_{2}\in[\beta_{2,\mathrm{min}},\beta_{2,\mathrm{max}}]$. This behavior could indicate a ``saturation" in the number of relevant natural states that can be built by exploring a collective degree of freedom and we could use this information to limit a priori the number of intrinsic states, i.e., before computing costly Hamiltonian overlaps and solving the HWG equation. This preselection can be performed by applying the following steps:

\begin{enumerate}[label=(\alph*)]
\item Find a sparse set of intrinsic wave functions in the interval $q_{i}\in[q_{\mathrm{min}},q_{\mathrm{max}}]$, $S=\lbrace|\Phi_{q_{i}}\rangle\rbrace_{i=1,...,N_{\mathrm{int}}}$
\item Solve Eq.~\eqref{norm_over_mat} and build the natural basis states (Eq.~\eqref{natural_basis}), $\lbrace|\Lambda_{k}\rangle\rbrace_{k=1,...,N_{\mathrm{int}}}$.
\item Build the $N_{\mathrm{int}}\times N_{\mathrm{int}}$ natural basis norm overlap matrix and keep only the $N_{\mathrm{nat}}$ states that fulfill the orthonormalization condition for a chosen $\varepsilon_{\mathrm{nat}}$ (Eq.~\eqref{orth_cond}).
\item If $N_{\mathrm{nat}}\simeq N_{\mathrm{int}}$, repeat the process but increasing the density of states in the set $S$ in order to reach the saturation point. Otherwise, solve the HWG equation (Eq.~\eqref{HWG_2}) by diagonalizing (without truncations related with $\varepsilon_{\lambda}$) the corresponding $N_{\mathrm{nat}}\times N_{\mathrm{nat}}$ Hamiltonian matrix.   
\end{enumerate}
Two a priori advantages of this protocol with respect to the plateau condition can be stressed. On the one hand, computing only norm overlap matrices and diagonal Hamiltonian overlaps is much less time-consuming than evaluating the full Hamiltonian overlap matrix. These are only calculated once the optimal number of elements in the natural basis is found. On the other hand, the method ensures the numerical orthonormality of the natural basis. Therefore, all the $N_{\mathrm{nat}}$ eigenvalues and eigenvectors obtained from the solution of the HWG equation should be meaningful in this space -not only those with the lowest energies- and the plateau condition should not be needed.   
\section{Results}\label{Results}
Let us now analyze the above protocol with actual PGCM calculations performed with the Gogny D1S energy density functional. Similar studies have been carried out with valence-space Hamiltonian PGCM methods using the recently developed code {\fontfamily{pcr}\selectfont
TAURUS
}~\cite{Bally19a,Bally21a,Sanchez_Fernandez21} and the results are equivalent and not shown for the sake of simplicity. Because of the versatility of the method, we can choose three nuclei as examples of double-magic ($^{40}$Ca), open-shell ($^{80}$Sr), and semi-magic ($^{186}$Pb) nuclei, all of them in different regions of the nuclear chart with different structural properties (see below and Refs.~\cite{Caurier07,Sienko03,Duguet03} and references therein). As mentioned above, the intrinsic wave functions that are mixed within the GCM ansatz are particle number and angular momentum projected (PNAMP) HFB quasiparticle vacua. These HFB states have different axial quadrupole deformations parametrized by $\beta_{2}$ and they are obtained using the VAPNP method, i.e., by minimizing the particle number projected energy with the constraint in the quadrupole moment operator, $\hat{Q}_{20}$. Additionally, we impose axial, parity, and simplex self-consistent symmetries to the underlying HFB transformation, and we use a spherical harmonic oscillator basis including thirteen major shells as the working basis to expand the HFB states (see Refs.~\cite{Egido16a,Robledo18a} for more details). Due to the self-consistent symmetries, the norm overlap matrix elements are calculated without ambiguities in their phases with the Onishi formula~\cite{Balian69a}.
\begin{figure*}[t]
\begin{center}
\includegraphics[width=1.0\textwidth]{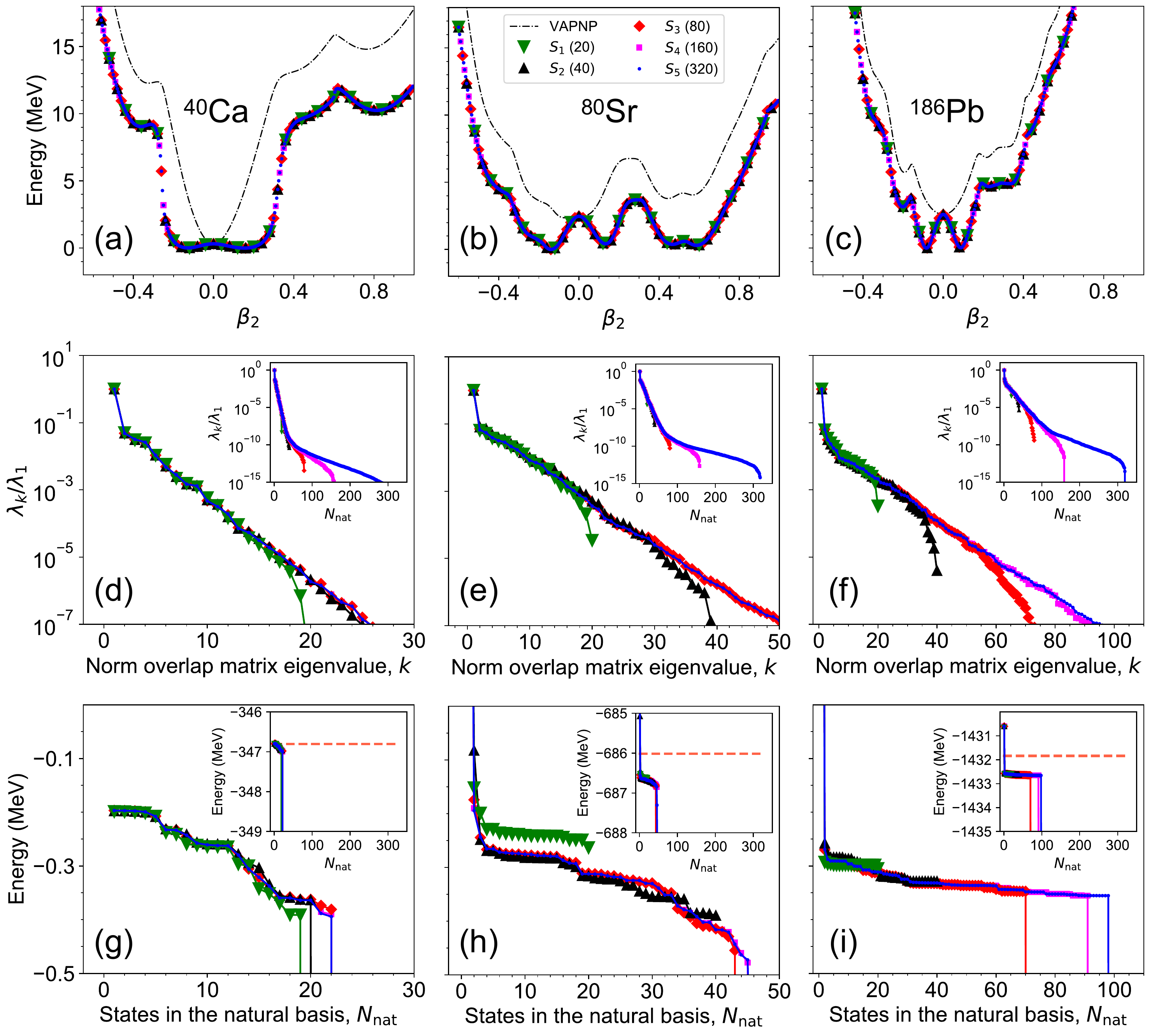}
\caption{(color online) (a)-(c) VAPNP (black dashed-dotted line) and PNAMP (symbols) total energies, normalized to their respective PNAMP minima, as a function of the axial quadrupole deformation calculated with the Gogny D1S interaction for different sets of initial states, $N_{\mathrm{int}}$, for the nuclei $^{40}$Ca (left panel), $^{80}$Sr (center panel) and $^{186}$Pb (right panel). (d)-(f) Eigenvalues of the norm overlap matrix sorted from the largest to the smallest values. (g)-(i) Lowest eigenvalues (shifted by 346.6, 686.4 and 1432.3 MeV) obtained by solving the HWG equations as a function of the number of the states included in the natural basis, $N_{\mathrm{nat}}$. The red dashed line corresponds to the minimum of the PNAMP energy obtained in (a)-(c) and the insets zoom out the same data represented in (d)-(f) and (g)-(i), respectively.}
\label{Fig1}
\end{center}
\end{figure*}

We will focus our analysis first on the ground state of the many-body system. Five different sets of intrinsic wave functions within the interval $\beta_{2}\in[-0.65,0.95]$ have been defined, namely, $S_{j=1,...,5}$ with $N_{\mathrm{int},j}=20,40,80,160,320$ elements in each set such that $S_{1}\subset S_{2}\subset ... \subset S_{5}$. In fact, we introduce one state between two consecutive states in the set $S_{j}$ to build the set $S_{j+1}$. The total energy curves (TECs) for the VAPNP and the subsequent PNAMP ($J=0$) approximations along the axial quadrupole deformation are represented in Fig.~\ref{Fig1}(a)-(c). The sets $S_{j}$ correspond to equally distributed points in such curves including always the points at the boundaries of the interval in $\beta_{2}$. It is important to stress that the choice of the interval in $\beta_{2}$ (and the points within it) is made to include the absolute minima of the different TECs and intrinsic states with VAPNP energies up to $\sim20$ MeV above such minima. The ability of the GCM to approach the exact solutions of the many-body system will obviously depend upon this selection, that is dictated by the diagonal norm and Hamiltonian overlaps. However, we will not focus in this work on the study of the GCM energies with respect to the limits of the interval and take the above prescription as a reasonable and widely used choice.    

We observe that, at the VAPNP level, the three nuclei show spherical minima with narrower wells for the magic ones ($^{40}$Ca and $^{186}$Pb) and some other local valleys at deformed configurations. Hence, an oblate normal-deformed and a prolate super-deformed minima are clearly seen in $^{40}$Ca, secondary prolate deformed minima are obtained in $^{80}$Sr, and prolate and oblate minima are also found in $^{186}$Pb. The simultaneous particle number and angular momentum restoration of the intrinsic HFB states ($J=0$) produces an energy gain in all the points of the VAPNP-TECs except the spherical one that becomes a maximum that separates two almost symmetric minima around this point. For $^{40}$Ca the PNAMP-TEC is rather flat within an interval of $\beta_{2}\in[-0.2,+0.2]$ and the normal-  and super-deformed minima gain a significant amount of correlation energy. The latter is also observed in $^{80}$Sr, where the PNAMP prolate deformed minimum is almost degenerated with the minima around the spherical point, and in the deformed VAPNP minima in $^{186}$Pb. Here, the two symmetric minima around $\beta_{2}=0$ are not as flat as in the $^{40}$Ca case. We note that the shape of the PNAMP-TECs are well reproduced independently of the density of points in the mesh, except in the nucleus $^{186}$Pb with the $S_{1}$ subset where the fine structure around the spherical and oblate minima could not be very well defined.
\subsection{Convergence of the GCM with the plateau condition}
Now we solve the HWG equations defined after mixing the $J=0$ intrinsic states for the different $S_{j}$ sets. The eigenvalues of the $N_{\mathrm{int},j}\times N_{\mathrm{int},j}$ norm overlap matrices, sorted from the largest to the smallest in the $x$-axis, are plotted in Fig.~\ref{Fig1}(d)-(f). These eigenvalues are normalized to the largest one. Insets represent the full range of eigenvalues obtained while a zoom-in to the most relevant regions are shown in the main body of the figures. We observe that the overall trends of the eigenvalues are independent of the nucleus under consideration. We also see common features in the behavior of the eigenvalues for all sets of intrinsic wave functions. Firstly, the largest eigenvalue is separated from the rest of the eigenvalues. Moreover, the larger the dimension of the norm overlap matrix is, the larger is the largest eigenvalue and smaller the smallest eigenvalue obtained (not shown because of the normalization). Secondly, the rest of eigenvalues decrease smoothly to very small values with nearly the same slope for all the sets in a rather wide range of eigenvalues. In fact, the curves of the normalized eigenvalues for the different sets are on top of each other until they deviate at the smaller eigenvalues obtained for each set. The differences between the sets are found in the change to a smaller slope for the $S_{5}$, $S_{4}$ and, to lesser extent, $S_{3}$ sets, that allows the accumulation of many eigenvalues in the range of $10^{-7}-10^{-12}$ (see insets). This behavior indicates that, even though the eigenvalue zero is an accumulation point, exact linear dependencies are not actually observed. In addition, potential problems associated to approximate linear dependencies will be more evident in the sets with more intrinsic HFB states.   

To check such potential numerical instabilities that could appear when solving the HWG equations we represent in Fig.~\ref{Fig1}(g)-(i) the lowest ($J=0$) eigenvalue of the $\langle\Lambda_{k}|\hat{H}|\Lambda_{k'}\rangle$ matrix (see Eq.~\eqref{HWG_2}) as a function of the number of states included in the natural basis, $N_{\mathrm{nat}}$, for the different sets, $S_{j}$. Again, the insets represent the same values at a lower-resolution scale. Most of the GCM points lie below the minimum of the PNAMP $J=0$ TEC shown in the top panel and marked as a dotted line in the insets at the bottom panel of the corresponding figures. This is an indication of the amount of correlations attained with the GCM mixing. 

We observe in $^{80}$Sr and $^{186}$Pb a strong reduction of the GCM energy from considering only one state in the natural basis (the one built with the largest eigenvalue of the norm) to adding more states. We also see that, after this reduction, the GCM energy is slowly decreasing with the number of states in the natural basis and sudden drops are obtained with the sets $S_{3,4,5}$ after considering a certain number of states, i.e., we have a sort of plateau conditions. For $^{40}$Ca we also see a similar result, but the initial jump is not found and the final abrupt drop is reached by all the sets. 

We point out that the plateaux obtained with the different sets lie almost on top of each other for the three nuclei considered, with the exception of the set $S_{1}$ in $^{80}$Sr that is slightly above. The main difference between the different sets is the larger size of the plateau observed for the sets with more intrinsic states. However, the size for $S_{3}$, $S_{4}$ and $S_{5}$ is significantly smaller than the number of intrinsic wave functions contained in the sets meaning that the approximate linear dependence is occurring (see insets). If we take the point where the first big jump in energy occurs to establish the maximum number of states in the natural basis allowed within the present intervals of deformation, $\beta_{2}$, then we obtain around 20 and 40 for $^{40}$Ca and $^{80}$Sr, respectively, independently of the initial set considered. For $^{186}$Pb, this value is $N_{\mathrm{nat}}=70$, $91$ and $98$ for the sets $S_{3}$, $S_{4}$ and $S_{5}$, and equal to the number of intrinsic states for the other two sets, $S_{1}$ and $S_{2}$. In fact, the last point before the first big jump in energy occurs in these nuclei for $\lambda_{k}/\lambda_{1}\approx10^{-7}$ (Fig.~\ref{Fig1}(d)-(f) and Fig.~\ref{Fig1}(g)-(i)). The usual choice of the final GCM energy (for a given set $S_{j}$) is to take such a last point before the approximate linear dependence is visible. From this point of view, the best PGCM energy is obtained with the $S_{5}$ set but the energy differences are all smaller than 50 keV except for $^{80}$Sr with the $S_{1}$ set that is $\sim 200$ keV. Nevertheless, the relative error is less than 1 per mil in the worst scenario. 
The above analysis suggests several conclusions, namely: a) we find a convergence of the PGCM ground state energy with the number of points included in a given interval of the collective coordinate; b) the normalized norm overlap matrix eigenvalues for the different sets lie on practically the same curve and the solutions of the HWG equations have a similar behavior; and, c) the number of numerically meaningful states in the natural basis seems to saturate because of the appearance of approximate linear dependencies. The latter is very significant because the construction of the natural basis does not require the evaluation of costly non-diagonal Hamiltonian overlaps.
\begin{figure*}[t]
\begin{center}
\includegraphics[width=\textwidth]{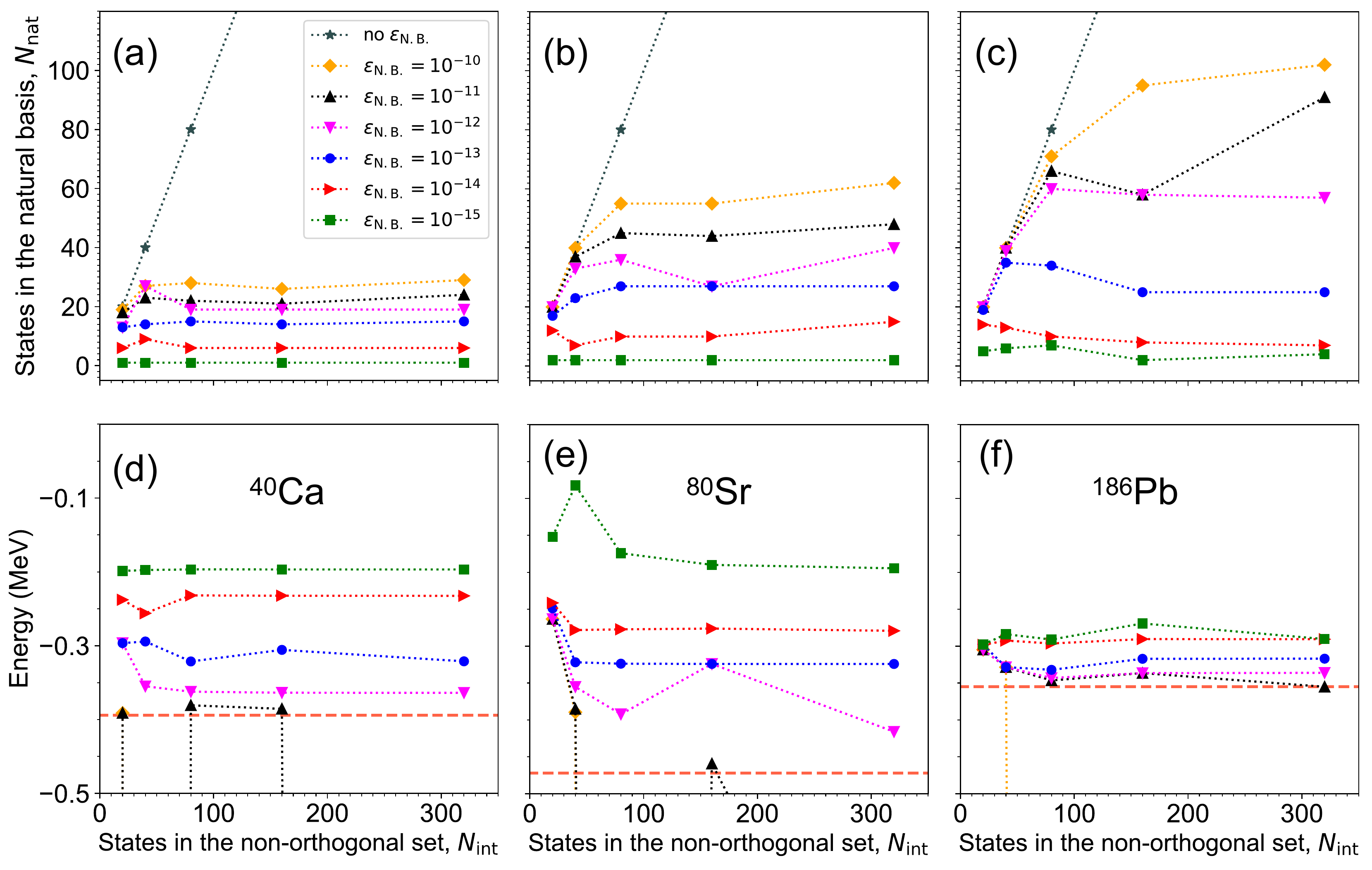}
\caption{(color online) (a)-(c) Number of states in the natural basis that fulfill Eq.~\eqref{orth_cond} for different values of $\varepsilon_{\mathrm{nat}}$ as a function of the number of intrinsic (non-orthonormal) states for the isotopes $^{40}$Ca (left panel), $^{80}$Sr (center panel) and $^{186}$Pb (right panel). (d)-(f) PGCM ground state energies computed with the natural bases represented in (a)-(c), shifted by 346.6, 686.4 and 1432.3 MeV, respectively. The dashed-red line corresponds to the lowest energy obtained before the big jump observed in Fig.~\ref{Fig1}(c).}
\label{Fig2}
\end{center}
\end{figure*}
\subsection{Convergence of the GCM with the orthonormality of the natural basis}
Let us now consider the condition of orthonormality of the states in the natural basis, Eq.~\eqref{orth_cond}, as a criterium to select numerically stable natural bases instead of the choice of a cutoff in the lowest eigenvalue of the norm overlap matrix. In Fig.~\ref{Fig2}(a)-(c) we represent for each set, $S_{j}$, the number of elements in the natural basis that fulfill such an orthonormality requirement defined by the parameter $\varepsilon_{\mathrm{nat}}$. We also vary such a value from no restriction at all to a precision of $\varepsilon_{\mathrm{nat}}=10^{-15}$. The overall behavior of the number of elements in the natural basis when increasing the number of intrinsic states is similar for the three isotopes. First of all, if no restriction in $\varepsilon_{\mathrm{nat}}$ is made, the natural basis coincides with the orthonormalization of the original set of intrinsic states. However, this large natural basis contains normally numerical instabilities that are manifested when the value of $\varepsilon_{\mathrm{nat}}$ decreases. Hence, for smaller $\varepsilon_{\mathrm{nat}}$ values, a reduction of the number of states in the natural basis is obtained. Such a reduction is more drastic for the denser $S_{4}$ and $S_{5}$ sets and for smaller values of $\varepsilon_{\mathrm{nat}}$. Moreover, for fixed $\varepsilon_{\mathrm{nat}}$ we obtain a saturation of $N_{\mathrm{nat}}$, i.e., including more intrinsic states in the original set does not produce larger natural bases. This is a key result of the present work because, based on this saturation property, one could determine, without evaluating costly non-diagonal Hamiltonian matrix elements, the minimum number of intrinsic states that maximizes the number of orthonormal states that can be built within a given interval of the generating coordinate. Such an optimal value would be given by, first, fixing $\varepsilon_{\mathrm{nat}}$ to a small value to ensure the orthonormality of the natural basis and, second, choose $N_{\mathrm{int}}$ in such a way that its corresponding natural basis fulfills the condition $N_{\mathrm{int}}\gtrsim N_{\mathrm{nat}}$. 

The above argument holds only in the case where natural bases with a similar number of elements but built from a different number of intrinsic states are equivalent when solving the HWG equation. To check such an assumption we represent in Fig.~\ref{Fig2}(d)-(e) the ground state energies obtained with the natural bases defined in the top panels of the same figures, i.e., for different combinations of $\varepsilon_{\mathrm{nat}}$ and $N_{\mathrm{int}}$. For $\varepsilon_{\mathrm{nat}}\leq10^{-12}$ we see that the PGCM energies are rather stable and have a very mild dependence with the number of intrinsic states included originally in the set. That means that we can reproduce the results obtained with many states in the initial set with much less states, showing the equivalency of the natural bases.
 
The PGCM energies obtained have a stronger dependence on $\varepsilon_{\mathrm{nat}}$. The number of states in the natural basis is generally larger for larger $\varepsilon_{\mathrm{nat}}$ and, consequently, the PGCM energies are smaller (the variational space is larger). However, for $\varepsilon_{\mathrm{nat}}\geq10^{-11}$ such small deviations from orthonormality of the natural basis introduce numerical instabilities, i.e., those natural bases still contain approximate linear dependences. Additionally, we see that the largest differences in the PGCM energies obtained with $\varepsilon_{\mathrm{nat}}=10^{-12}$ and $10^{-15}$ (stable cases) are $170$ keV, $220$ keV and $45$ keV for $^{40}$Ca, $^{80}$Sr and $^{186}$Pb, respectively. As a reference for the PGCM energy, we also plot in Fig.~\ref{Fig2}(d)-(f) the PGCM energy obtained from the plateau condition discussed previously. Even though these values are smaller to those obtained with $\varepsilon_{\mathrm{nat}}$ as a convergence criterium (instead of $\varepsilon_{\lambda}$), the energy differences are not larger than 200 keV. Therefore, the value of $\varepsilon_{\mathrm{nat}}$ should be small enough to ensure a numerically sound natural basis but not too small to leave natural bases made of too few states. For this set of nuclei and EDFs, $\varepsilon_{\mathrm{nat}}\approx10^{-12}-10^{-13}$ seem to be reasonable choices.  

\begin{figure}[t]
\begin{center}
\includegraphics[width=\columnwidth]{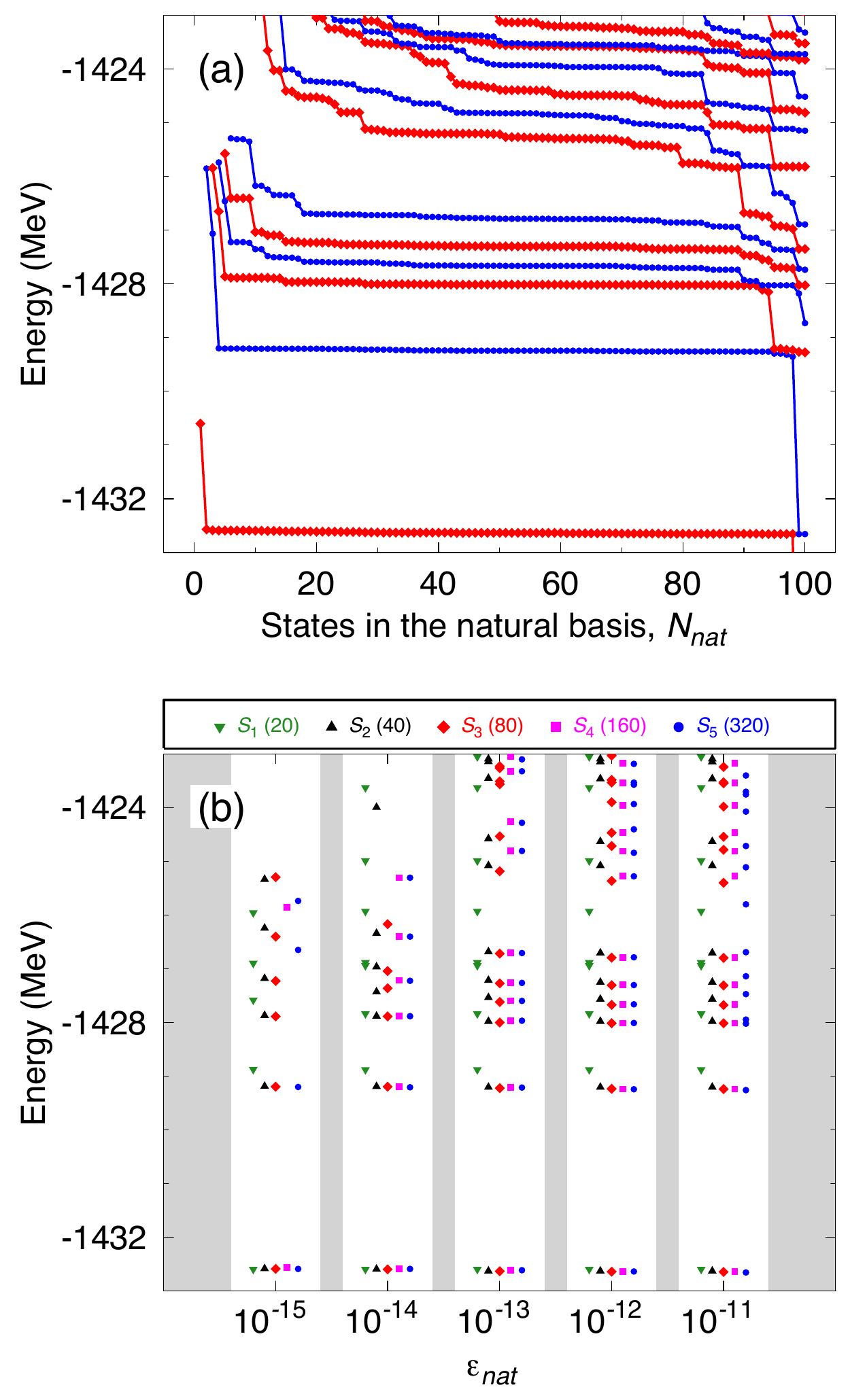}
\caption{(color online) (a) Energies of the $0^{+}$ states (up to 10 MeV of excitation energies) obtained by solving the HWG equations as a function of the number of the states included in the natural basis, $N_{\mathrm{nat}}$ for the set $S_{5}$ ($N_{\mathrm{int}}=320$) and the nucleus $^{186}$Pb. Red (blue) filled diamonds (bullets) represent the energies $E^{\sigma}$ with $\sigma=1,3,5,...$ ($\sigma=2,4,6,...$) and are connected with lines to guide the eye. (b) Energies of the $0^{+}$ states (up to 10 MeV of excitation energies) obtained by solving the HWG equations for the different sets, $S_{j}=1-5$, and six values for the orthonormality condition of the natural basis, $\varepsilon_{\mathrm{nat}}=10^{-15}, 10^{-14}, 10^{-13}, 10^{-12},10^{-11}$, for the same nucleus ($^{186}$Pb).}
\label{Fig3}
\end{center}
\end{figure}
\subsection{Convergence of the GCM excited states}
We finally discuss the convergence and meaningfulness of the excited $E^{\sigma}$ states obtained with the PGCM method. The standard procedure to determine the energies of the excited states is similar to the one used in the ground state. The HWG equation is solved for a given set of intrinsic wave functions and the resulting energies are studied as a function of the number of states in the natural basis. Such a basis is sorted out by introducing first the states in the natural basis that are built with the eigenvectors (and the eigenvalues) of the norm overlap matrix with the largest eigenvalues. In Fig.~\ref{Fig3}(a) these energies ($J=0$) are represented in a range around 10 MeV of excitation energy for the set $S_{5}$ and the nucleus $^{186}$Pb. The values of $\sigma=1-16$ are shown within this range. Here we can clearly see that, as soon as $\sigma$ is increased, the plateaux are worse defined and show jumps much earlier than the appearance of the approximate linear dependence in the lowest eigenvalue (at $N_{\mathrm{nat}}=98$ in this case). Some of these discontinuities fall down to the next lower plateau due to the appearance of one state with lower energy that makes that such a state, formerly located at a given $\sigma$ for a small $N_{\mathrm{nat}}$, is now found at $\sigma+1$ or even larger for larger $N_{\mathrm{nat}}$. These jumps are most likely a consequence of the approximate linear dependence and many of the states that are obtained in the diagonalization of the Hamiltonian matrix defined with a large value of $N_{\mathrm{nat}}$ are spurious. For example, the first two excited $0^{+}$ states with $\sigma=2,3$ obtained with $N_{\mathrm{nat}}=98$ (the position of the big jump in the ground state energy) are almost (and suspiciously) degenerated. This result differs from the one obtained if a smaller value for $N_{\mathrm{nat}}$ is chosen (or, equivalently, a larger value for $\varepsilon_{\lambda}$). In such a case, clearly distinguishable plateaux for $\sigma=1-6$ are obtained within the same range of $N_{\mathrm{nat}}$, e.g., $N_{\mathrm{nat}}=30-70$. Then, computing the values of the excitation energies with the natural basis defined just before the linear dependence explosion observed in the ground state could be meaningless~\footnote{A way to determine the energies for all values of $\sigma$ that could be explored in the future within the plateau condition could be the study of the extended plateaux at large $N_{\mathrm{nat}}$ that are formed by joining the points of different $\sigma$'s that show certain continuity~\cite{EgidoPriv}.}.

Let us discuss now, for the same nucleus, the results obtained whenever we use the orthonormality of the natural basis as the criterium to select the number of states included in it. 
Calculations are performed by selecting a set of intrinsic wave functions, $S_{j=1-5}$, and a given value for the orthonormality condition, $\varepsilon_{\mathrm{nat}}$. Hence, the HWG equation is solved with the maximum number of states in the natural basis allowed by $\varepsilon_{\mathrm{nat}}$. The results (the eigenvalues of the Hamiltonian matrix written in such a natural basis) are shown in Fig.~\ref{Fig3}(b) within the same range of energies as in Fig.~\ref{Fig3}(a). Here we observe again that the states with $\sigma=1$ are well converged for the range of $\varepsilon_{\mathrm{nat}}$ values plotted in the figure and their energy is practically independent of the initial set, $S_{j=1-5}$. Moreover, for smaller values of $\varepsilon_{\mathrm{nat}}$, i.e., more restricted orthonormality of the natural basis, the number of excited states is smaller (see also Fig.~\ref{Fig2}(c)). For $\varepsilon_{\mathrm{nat}}=10^{-15}$ few and rather dependent on the initial set excited states are obtained. For the rest of values of $\varepsilon_{\mathrm{nat}}$ we observe that the set with a smaller number of intrinsic states, $S_{1}$, is the one that deviates most from the rest, even for small values of $\sigma$. However, the results show an excellent agreement for the different initial sets $S_{j=2-5}$ in the states with $\sigma=2-6$ for values of $\varepsilon_{\mathrm{nat}}=10^{-13}-10^{-12}$. For larger $\sigma$, the energies are rather dependent both on $\varepsilon_{\mathrm{nat}}$ and $S_{j}$ and some degenerated states are also observed, e.g., the states with $\sigma=3,4$ for $\varepsilon_{\mathrm{nat}}=10^{-11}$ and the set $S_{5}$. Therefore, we can conclude from this case that the $\sigma\neq1$ excited states show larger dependencies than the $\sigma=1$ state both on the parameter that controls the orthonormality of the natural basis and the number of intrinsic states. These dependencies are larger for larger values of $\sigma$ and smaller values of $\varepsilon_{\mathrm{nat}}$. These uncertainties can be a limiting factor to use the PGCM method to study, e.g., excited states at high energy or level densities. On the positive side, we can also assure that the results obtained with a very dense initial set of intrinsic wave functions can be reproduced with initial sets containing much less intrinsic wave functions that are determined by the orthonormality condition, i.e., before computing costly Hamiltonian kernels. In the present case, if we choose $\varepsilon_{\mathrm{nat}}=10^{-13}$ and follow the protocol proposed above, the set that fulfills $N_{\mathrm{int}} \gtrsim N_{\mathrm{nat}}$ with the minimum number of $N_{\mathrm{int}}$ is $S_{2}$ ($N_{\mathrm{int}}=40$). We see in Fig.~\ref{Fig3}(b) that the results obtained in this case for $\sigma=1-6$ are very much alike to those obtained with sets made of many more intrinsic states. 
\section{Summary and conclusions}\label{Summary}
To summarize, we have analyzed the determination of PGCM energies once a collective coordinate and the interval in which such a coordinate is defined are chosen. In our case, Gogny EDF calculations calculations of three selected nuclei have been performed using the axial quadrupole degree of freedom as the generating coordinate and including particle number and angular momentum symmetry restorations. Several sets of initial intrinsic states have been defined to study the convergence of the results with the number of initial states -within the same interval. These states are not orthonormal and can contain (exact or approximate) linear dependencies.
We have observed that the largest normalized eigenvalues of the norm overlap matrices are very similar for all the sets although the sets with more intrinsic states accumulate more small eigenvalues. We also see similar energy plateaux for the different sets suggesting that a lot of redundancies (or approximate linear dependencies) are contained in the sets with a larger number of intrinsic states. 

Inspired by these results, we have proposed a way of determining a priori (without evaluating Hamiltonian overlaps) natural bases with much less approximate linear dependencies based on their orthonormality conditions. We have clearly seen the saturation of the number of states in the natural basis with the number of initial states. Moreover, the results obtained with a similar number of states in the natural basis that are coming from a very much different number of states in the initial sets are strikingly similar.

Finally, we have also analyzed the excited states ($\sigma\neq1$) and we have seen the difficulties of using the plateau criterium to determine the excitation energies reliably. Using the orthonormality of the natural basis as a criterium reduces the uncertainties in the lowest values of $\sigma$ but the situation is still out of control for higher excitation energies. Nevertheless, the main outcome of this work is that one could define, before computing expensive non-diagonal Hamiltonian kernels, an initial set of non-orthogonal states containing most of the physical information but a small amount (or none) of approximate linear dependencies. Applications to other Hamiltonians, multiple degrees of freedom and other observables beyond the $0^{+}$ energies are left over for future analyses.

\section*{Acknowledgements}

We would like to specially thank B. Bally for his valuable comments and discussions. TRR also thanks J. L. Egido, J. Engel, F. Nowacki, L. M. Robledo, A. M. Romero, A. S\'anchez-Fern\'andez and J. M. Yao for useful discussions. This work was supported by the Spanish MICINN under PGC2018-094583-B-I00 and PRE2019-088036. We gratefully thank the support from GSI-Darmstadt computing facility.


\bibliography{biblio.bib}

\begin{thebibliography}{32}%
\makeatletter
\providecommand \@ifxundefined [1]{%
 \@ifx{#1\undefined}
}%
\providecommand \@ifnum [1]{%
 \ifnum #1\expandafter \@firstoftwo
 \else \expandafter \@secondoftwo
 \fi
}%
\providecommand \@ifx [1]{%
 \ifx #1\expandafter \@firstoftwo
 \else \expandafter \@secondoftwo
 \fi
}%
\providecommand \natexlab [1]{#1}%
\providecommand \enquote  [1]{``#1''}%
\providecommand \bibnamefont  [1]{#1}%
\providecommand \bibfnamefont [1]{#1}%
\providecommand \citenamefont [1]{#1}%
\providecommand \href@noop [0]{\@secondoftwo}%
\providecommand \href [0]{\begingroup \@sanitize@url \@href}%
\providecommand \@href[1]{\@@startlink{#1}\@@href}%
\providecommand \@@href[1]{\endgroup#1\@@endlink}%
\providecommand \@sanitize@url [0]{\catcode `\\12\catcode `\$12\catcode
  `\&12\catcode `\#12\catcode `\^12\catcode `\_12\catcode `\%12\relax}%
\providecommand \@@startlink[1]{}%
\providecommand \@@endlink[0]{}%
\providecommand \url  [0]{\begingroup\@sanitize@url \@url }%
\providecommand \@url [1]{\endgroup\@href {#1}{\urlprefix }}%
\providecommand \urlprefix  [0]{URL }%
\providecommand \Eprint [0]{\href }%
\providecommand \doibase [0]{http://dx.doi.org/}%
\providecommand \selectlanguage [0]{\@gobble}%
\providecommand \bibinfo  [0]{\@secondoftwo}%
\providecommand \bibfield  [0]{\@secondoftwo}%
\providecommand \translation [1]{[#1]}%
\providecommand \BibitemOpen [0]{}%
\providecommand \bibitemStop [0]{}%
\providecommand \bibitemNoStop [0]{.\EOS\space}%
\providecommand \EOS [0]{\spacefactor3000\relax}%
\providecommand \BibitemShut  [1]{\csname bibitem#1\endcsname}%
\let\auto@bib@innerbib\@empty
\bibitem [{\citenamefont {Hill}\ and\ \citenamefont {Wheeler}(1953)}]{Hill53a}%
  \BibitemOpen
  \bibfield  {author} {\bibinfo {author} {\bibfnamefont {D.~L.}\ \bibnamefont
  {Hill}}\ and\ \bibinfo {author} {\bibfnamefont {J.~A.}\ \bibnamefont
  {Wheeler}},\ }\href {\doibase 10.1103/PhysRev.89.1102} {\bibfield  {journal}
  {\bibinfo  {journal} {Phys. Rev.}\ }\textbf {\bibinfo {volume} {89}},\
  \bibinfo {pages} {1102} (\bibinfo {year} {1953})}\BibitemShut {NoStop}%
\bibitem [{\citenamefont {Griffin}\ and\ \citenamefont
  {Wheeler}(1957)}]{Griffin57a}%
  \BibitemOpen
  \bibfield  {author} {\bibinfo {author} {\bibfnamefont {J.~J.}\ \bibnamefont
  {Griffin}}\ and\ \bibinfo {author} {\bibfnamefont {J.~A.}\ \bibnamefont
  {Wheeler}},\ }\href {\doibase 10.1103/PhysRev.108.311} {\bibfield  {journal}
  {\bibinfo  {journal} {Phys. Rev.}\ }\textbf {\bibinfo {volume} {108}},\
  \bibinfo {pages} {311} (\bibinfo {year} {1957})}\BibitemShut {NoStop}%
\bibitem [{\citenamefont {Ring}\ and\ \citenamefont {Schuck}(1980)}]{RS80a}%
  \BibitemOpen
  \bibfield  {author} {\bibinfo {author} {\bibfnamefont {P.}~\bibnamefont
  {Ring}}\ and\ \bibinfo {author} {\bibfnamefont {P.}~\bibnamefont {Schuck}},\
  }\href@noop {} {\emph {\bibinfo {title} {The Nuclear Many-Body Problem}}}\
  (\bibinfo  {publisher} {Springer-Verlag},\ \bibinfo {address} {New York},\
  \bibinfo {year} {1980})\BibitemShut {NoStop}%
\bibitem [{\citenamefont {Bender}\ \emph {et~al.}(2003)\citenamefont {Bender},
  \citenamefont {Heenen},\ and\ \citenamefont {Reinhard}}]{Bender03a}%
  \BibitemOpen
  \bibfield  {author} {\bibinfo {author} {\bibfnamefont {M.}~\bibnamefont
  {Bender}}, \bibinfo {author} {\bibfnamefont {P.-H.}\ \bibnamefont {Heenen}},
  \ and\ \bibinfo {author} {\bibfnamefont {P.-G.}\ \bibnamefont {Reinhard}},\
  }\href {\doibase 10.1103/RevModPhys.75.121} {\bibfield  {journal} {\bibinfo
  {journal} {Rev. Mod. Phys.}\ }\textbf {\bibinfo {volume} {75}},\ \bibinfo
  {pages} {121} (\bibinfo {year} {2003})}\BibitemShut {NoStop}%
\bibitem [{\citenamefont {Lacroix}\ \emph {et~al.}(2009)\citenamefont
  {Lacroix}, \citenamefont {Duguet},\ and\ \citenamefont
  {Bender}}]{Lacroix09a}%
  \BibitemOpen
  \bibfield  {author} {\bibinfo {author} {\bibfnamefont {D.}~\bibnamefont
  {Lacroix}}, \bibinfo {author} {\bibfnamefont {T.}~\bibnamefont {Duguet}}, \
  and\ \bibinfo {author} {\bibfnamefont {M.}~\bibnamefont {Bender}},\ }\href
  {\doibase 10.1103/PhysRevC.79.044318} {\bibfield  {journal} {\bibinfo
  {journal} {Phys. Rev. C}\ }\textbf {\bibinfo {volume} {79}},\ \bibinfo
  {pages} {044318} (\bibinfo {year} {2009})}\BibitemShut {NoStop}%
\bibitem [{\citenamefont {Nikšić}\ \emph {et~al.}(2011)\citenamefont
  {Nikšić}, \citenamefont {Vretenar},\ and\ \citenamefont
  {Ring}}]{Niksic11a}%
  \BibitemOpen
  \bibfield  {author} {\bibinfo {author} {\bibfnamefont {T.}~\bibnamefont
  {Nikšić}}, \bibinfo {author} {\bibfnamefont {D.}~\bibnamefont {Vretenar}},
  \ and\ \bibinfo {author} {\bibfnamefont {P.}~\bibnamefont {Ring}},\ }\href
  {\doibase https://doi.org/10.1016/j.ppnp.2011.01.055} {\bibfield  {journal}
  {\bibinfo  {journal} {Progress in Particle and Nuclear Physics}\ }\textbf
  {\bibinfo {volume} {66}},\ \bibinfo {pages} {519 } (\bibinfo {year}
  {2011})}\BibitemShut {NoStop}%
\bibitem [{\citenamefont {Egido}(2016)}]{Egido16a}%
  \BibitemOpen
  \bibfield  {author} {\bibinfo {author} {\bibfnamefont {J.~L.}\ \bibnamefont
  {Egido}},\ }\href {http://stacks.iop.org/1402-4896/91/i=7/a=073003}
  {\bibfield  {journal} {\bibinfo  {journal} {Physica Scripta}\ }\textbf
  {\bibinfo {volume} {91}},\ \bibinfo {pages} {073003} (\bibinfo {year}
  {2016})}\BibitemShut {NoStop}%
\bibitem [{\citenamefont {Robledo}\ \emph {et~al.}(2018)\citenamefont
  {Robledo}, \citenamefont {Rodr{\'{\i}}guez},\ and\ \citenamefont
  {Rodr{\'{\i}}guez-Guzm{\'{a}}n}}]{Robledo18a}%
  \BibitemOpen
  \bibfield  {author} {\bibinfo {author} {\bibfnamefont {L.~M.}\ \bibnamefont
  {Robledo}}, \bibinfo {author} {\bibfnamefont {T.~R.}\ \bibnamefont
  {Rodr{\'{\i}}guez}}, \ and\ \bibinfo {author} {\bibfnamefont {R.~R.}\
  \bibnamefont {Rodr{\'{\i}}guez-Guzm{\'{a}}n}},\ }\href {\doibase
  10.1088/1361-6471/aadebd} {\bibfield  {journal} {\bibinfo  {journal} {Journal
  of Physics G: Nuclear and Particle Physics}\ }\textbf {\bibinfo {volume}
  {46}},\ \bibinfo {pages} {013001} (\bibinfo {year} {2018})}\BibitemShut
  {NoStop}%
\bibitem [{\citenamefont {Jiao}\ \emph {et~al.}(2017)\citenamefont {Jiao},
  \citenamefont {Engel},\ and\ \citenamefont {Holt}}]{Jiao17a}%
  \BibitemOpen
  \bibfield  {author} {\bibinfo {author} {\bibfnamefont {C.~F.}\ \bibnamefont
  {Jiao}}, \bibinfo {author} {\bibfnamefont {J.}~\bibnamefont {Engel}}, \ and\
  \bibinfo {author} {\bibfnamefont {J.~D.}\ \bibnamefont {Holt}},\ }\href
  {\doibase 10.1103/PhysRevC.96.054310} {\bibfield  {journal} {\bibinfo
  {journal} {Phys. Rev. C}\ }\textbf {\bibinfo {volume} {96}},\ \bibinfo
  {pages} {054310} (\bibinfo {year} {2017})}\BibitemShut {NoStop}%
\bibitem [{\citenamefont {Bally}\ \emph {et~al.}(2019)\citenamefont {Bally},
  \citenamefont {S\'anchez-Fern\'andez},\ and\ \citenamefont
  {Rodr\'{\i}guez}}]{Bally19a}%
  \BibitemOpen
  \bibfield  {author} {\bibinfo {author} {\bibfnamefont {B.}~\bibnamefont
  {Bally}}, \bibinfo {author} {\bibfnamefont {A.}~\bibnamefont
  {S\'anchez-Fern\'andez}}, \ and\ \bibinfo {author} {\bibfnamefont {T.~R.}\
  \bibnamefont {Rodr\'{\i}guez}},\ }\href {\doibase
  10.1103/PhysRevC.100.044308} {\bibfield  {journal} {\bibinfo  {journal}
  {Phys. Rev. C}\ }\textbf {\bibinfo {volume} {100}},\ \bibinfo {pages}
  {044308} (\bibinfo {year} {2019})}\BibitemShut {NoStop}%
\bibitem [{\citenamefont {S\'anchez-Fern\'andez}\ \emph
  {et~al.}(2021)\citenamefont {S\'anchez-Fern\'andez}, \citenamefont {Bally},\
  and\ \citenamefont {Rodr\'{\i}guez}}]{Sanchez_Fernandez21}%
  \BibitemOpen
  \bibfield  {author} {\bibinfo {author} {\bibfnamefont {A.}~\bibnamefont
  {S\'anchez-Fern\'andez}}, \bibinfo {author} {\bibfnamefont {B.}~\bibnamefont
  {Bally}}, \ and\ \bibinfo {author} {\bibfnamefont {T.~R.}\ \bibnamefont
  {Rodr\'{\i}guez}},\ }\href {\doibase 10.1103/PhysRevC.104.054306} {\bibfield
  {journal} {\bibinfo  {journal} {Phys. Rev. C}\ }\textbf {\bibinfo {volume}
  {104}},\ \bibinfo {pages} {054306} (\bibinfo {year} {2021})}\BibitemShut
  {NoStop}%
\bibitem [{\citenamefont {Frosini}\ \emph
  {et~al.}(2022{\natexlab{a}})\citenamefont {Frosini}, \citenamefont {Duguet},
  \citenamefont {Ebran}, \citenamefont {Bally}, \citenamefont {Mongelli},
  \citenamefont {Rodr{\'i}guez}, \citenamefont {Roth},\ and\ \citenamefont
  {Som{\`a}}}]{Frosini22b}%
  \BibitemOpen
  \bibfield  {author} {\bibinfo {author} {\bibfnamefont {M.}~\bibnamefont
  {Frosini}}, \bibinfo {author} {\bibfnamefont {T.}~\bibnamefont {Duguet}},
  \bibinfo {author} {\bibfnamefont {J.-P.}\ \bibnamefont {Ebran}}, \bibinfo
  {author} {\bibfnamefont {B.}~\bibnamefont {Bally}}, \bibinfo {author}
  {\bibfnamefont {T.}~\bibnamefont {Mongelli}}, \bibinfo {author}
  {\bibfnamefont {T.~R.}\ \bibnamefont {Rodr{\'i}guez}}, \bibinfo {author}
  {\bibfnamefont {R.}~\bibnamefont {Roth}}, \ and\ \bibinfo {author}
  {\bibfnamefont {V.}~\bibnamefont {Som{\`a}}},\ }\href {\doibase
  10.1140/epja/s10050-022-00693-y} {\bibfield  {journal} {\bibinfo  {journal}
  {The European Physical Journal A}\ }\textbf {\bibinfo {volume} {58}},\
  \bibinfo {pages} {63} (\bibinfo {year} {2022}{\natexlab{a}})}\BibitemShut
  {NoStop}%
\bibitem [{\citenamefont {Frosini}\ \emph
  {et~al.}(2022{\natexlab{b}})\citenamefont {Frosini}, \citenamefont {Duguet},
  \citenamefont {Ebran}, \citenamefont {Bally}, \citenamefont {Hergert},
  \citenamefont {Rodr{\'i}guez}, \citenamefont {Roth}, \citenamefont {Yao},\
  and\ \citenamefont {Som{\`a}}}]{Frosini22c}%
  \BibitemOpen
  \bibfield  {author} {\bibinfo {author} {\bibfnamefont {M.}~\bibnamefont
  {Frosini}}, \bibinfo {author} {\bibfnamefont {T.}~\bibnamefont {Duguet}},
  \bibinfo {author} {\bibfnamefont {J.-P.}\ \bibnamefont {Ebran}}, \bibinfo
  {author} {\bibfnamefont {B.}~\bibnamefont {Bally}}, \bibinfo {author}
  {\bibfnamefont {H.}~\bibnamefont {Hergert}}, \bibinfo {author} {\bibfnamefont
  {T.~R.}\ \bibnamefont {Rodr{\'i}guez}}, \bibinfo {author} {\bibfnamefont
  {R.}~\bibnamefont {Roth}}, \bibinfo {author} {\bibfnamefont {J.~M.}\
  \bibnamefont {Yao}}, \ and\ \bibinfo {author} {\bibfnamefont
  {V.}~\bibnamefont {Som{\`a}}},\ }\href {\doibase
  10.1140/epja/s10050-022-00694-x} {\bibfield  {journal} {\bibinfo  {journal}
  {The European Physical Journal A}\ }\textbf {\bibinfo {volume} {58}},\
  \bibinfo {pages} {64} (\bibinfo {year} {2022}{\natexlab{b}})}\BibitemShut
  {NoStop}%
\bibitem [{\citenamefont {Dao}\ and\ \citenamefont {Nowacki}(2022)}]{Dao22}%
  \BibitemOpen
  \bibfield  {author} {\bibinfo {author} {\bibfnamefont {D.~D.}\ \bibnamefont
  {Dao}}\ and\ \bibinfo {author} {\bibfnamefont {F.}~\bibnamefont {Nowacki}},\
  }\href {\doibase 10.1103/PhysRevC.105.054314} {\bibfield  {journal} {\bibinfo
   {journal} {Phys. Rev. C}\ }\textbf {\bibinfo {volume} {105}},\ \bibinfo
  {pages} {054314} (\bibinfo {year} {2022})}\BibitemShut {NoStop}%
\bibitem [{\citenamefont {Shimizu}\ \emph {et~al.}(2021)\citenamefont
  {Shimizu}, \citenamefont {Mizusaki}, \citenamefont {Kaneko},\ and\
  \citenamefont {Tsunoda}}]{Shimizu21a}%
  \BibitemOpen
  \bibfield  {author} {\bibinfo {author} {\bibfnamefont {N.}~\bibnamefont
  {Shimizu}}, \bibinfo {author} {\bibfnamefont {T.}~\bibnamefont {Mizusaki}},
  \bibinfo {author} {\bibfnamefont {K.}~\bibnamefont {Kaneko}}, \ and\ \bibinfo
  {author} {\bibfnamefont {Y.}~\bibnamefont {Tsunoda}},\ }\href {\doibase
  10.1103/PhysRevC.103.064302} {\bibfield  {journal} {\bibinfo  {journal}
  {Phys. Rev. C}\ }\textbf {\bibinfo {volume} {103}},\ \bibinfo {pages}
  {064302} (\bibinfo {year} {2021})}\BibitemShut {NoStop}%
\bibitem [{\citenamefont {Romero}\ \emph {et~al.}(2021)\citenamefont {Romero},
  \citenamefont {Yao}, \citenamefont {Bally}, \citenamefont {Rodr\'{\i}guez},\
  and\ \citenamefont {Engel}}]{Romero21}%
  \BibitemOpen
  \bibfield  {author} {\bibinfo {author} {\bibfnamefont {A.~M.}\ \bibnamefont
  {Romero}}, \bibinfo {author} {\bibfnamefont {J.~M.}\ \bibnamefont {Yao}},
  \bibinfo {author} {\bibfnamefont {B.}~\bibnamefont {Bally}}, \bibinfo
  {author} {\bibfnamefont {T.~R.}\ \bibnamefont {Rodr\'{\i}guez}}, \ and\
  \bibinfo {author} {\bibfnamefont {J.}~\bibnamefont {Engel}},\ }\href
  {\doibase 10.1103/PhysRevC.104.054317} {\bibfield  {journal} {\bibinfo
  {journal} {Phys. Rev. C}\ }\textbf {\bibinfo {volume} {104}},\ \bibinfo
  {pages} {054317} (\bibinfo {year} {2021})}\BibitemShut {NoStop}%
\bibitem [{\citenamefont {Broeckhove}\ and\ \citenamefont
  {Deumens}(1979)}]{Broeckhove1979}%
  \BibitemOpen
  \bibfield  {author} {\bibinfo {author} {\bibfnamefont {J.}~\bibnamefont
  {Broeckhove}}\ and\ \bibinfo {author} {\bibfnamefont {E.}~\bibnamefont
  {Deumens}},\ }\href {\doibase 10.1007/BF01547468} {\bibfield  {journal}
  {\bibinfo  {journal} {Zeitschrift f{\"u}r Physik A Atoms and Nuclei}\
  }\textbf {\bibinfo {volume} {292}},\ \bibinfo {pages} {243} (\bibinfo {year}
  {1979})}\BibitemShut {NoStop}%
\bibitem [{\citenamefont {L\"owdin}(1956)}]{Lowdin1956}%
  \BibitemOpen
  \bibfield  {author} {\bibinfo {author} {\bibfnamefont {P.-O.}\ \bibnamefont
  {L\"owdin}},\ }\href {\doibase 10.1080/00018735600101155} {\bibfield
  {journal} {\bibinfo  {journal} {Advances in Physics}\ }\textbf {\bibinfo
  {volume} {5}},\ \bibinfo {pages} {1} (\bibinfo {year} {1956})},\ \Eprint
  {http://arxiv.org/abs/https://doi.org/10.1080/00018735600101155}
  {https://doi.org/10.1080/00018735600101155} \BibitemShut {NoStop}%
\bibitem [{\citenamefont {Lathouwers}(1976{\natexlab{a}})}]{Lathouwers1976a}%
  \BibitemOpen
  \bibfield  {author} {\bibinfo {author} {\bibfnamefont {L.}~\bibnamefont
  {Lathouwers}},\ }\href {\doibase
  https://doi.org/10.1016/0003-4916(76)90171-8} {\bibfield  {journal} {\bibinfo
   {journal} {Annals of Physics}\ }\textbf {\bibinfo {volume} {102}},\ \bibinfo
  {pages} {347} (\bibinfo {year} {1976}{\natexlab{a}})}\BibitemShut {NoStop}%
\bibitem [{\citenamefont {de~Toledo~Piza}\ \emph {et~al.}(1977)\citenamefont
  {de~Toledo~Piza}, \citenamefont {de~Passos}, \citenamefont {Galetti},
  \citenamefont {Nemes},\ and\ \citenamefont {Watanabe}}]{Piza1977}%
  \BibitemOpen
  \bibfield  {author} {\bibinfo {author} {\bibfnamefont {A.~F.~R.}\
  \bibnamefont {de~Toledo~Piza}}, \bibinfo {author} {\bibfnamefont {E.~J.~V.}\
  \bibnamefont {de~Passos}}, \bibinfo {author} {\bibfnamefont {D.}~\bibnamefont
  {Galetti}}, \bibinfo {author} {\bibfnamefont {M.~C.}\ \bibnamefont {Nemes}},
  \ and\ \bibinfo {author} {\bibfnamefont {M.~M.}\ \bibnamefont {Watanabe}},\
  }\href {\doibase 10.1103/PhysRevC.15.1477} {\bibfield  {journal} {\bibinfo
  {journal} {Phys. Rev. C}\ }\textbf {\bibinfo {volume} {15}},\ \bibinfo
  {pages} {1477} (\bibinfo {year} {1977})}\BibitemShut {NoStop}%
\bibitem [{\citenamefont {Bonche}\ \emph {et~al.}(1990)\citenamefont {Bonche},
  \citenamefont {Dobaczewski}, \citenamefont {Flocard}, \citenamefont
  {Heenen},\ and\ \citenamefont {Meyer}}]{Bonche90a}%
  \BibitemOpen
  \bibfield  {author} {\bibinfo {author} {\bibfnamefont {P.}~\bibnamefont
  {Bonche}}, \bibinfo {author} {\bibfnamefont {J.}~\bibnamefont {Dobaczewski}},
  \bibinfo {author} {\bibfnamefont {H.}~\bibnamefont {Flocard}}, \bibinfo
  {author} {\bibfnamefont {P.-H.}\ \bibnamefont {Heenen}}, \ and\ \bibinfo
  {author} {\bibfnamefont {J.}~\bibnamefont {Meyer}},\ }\href {\doibase
  https://doi.org/10.1016/0375-9474(90)90062-Q} {\bibfield  {journal} {\bibinfo
   {journal} {Nuclear Physics A}\ }\textbf {\bibinfo {volume} {510}},\ \bibinfo
  {pages} {466 } (\bibinfo {year} {1990})}\BibitemShut {NoStop}%
\bibitem [{\citenamefont {Lathouwers}(1976{\natexlab{b}})}]{Lathouwers1976b}%
  \BibitemOpen
  \bibfield  {author} {\bibinfo {author} {\bibfnamefont {L.}~\bibnamefont
  {Lathouwers}},\ }\href {\doibase https://doi.org/10.1002/qua.560100304}
  {\bibfield  {journal} {\bibinfo  {journal} {International Journal of Quantum
  Chemistry}\ }\textbf {\bibinfo {volume} {10}},\ \bibinfo {pages} {413}
  (\bibinfo {year} {1976}{\natexlab{b}})},\ \Eprint
  {http://arxiv.org/abs/https://onlinelibrary.wiley.com/doi/pdf/10.1002/qua.560100304}
  {https://onlinelibrary.wiley.com/doi/pdf/10.1002/qua.560100304} \BibitemShut
  {NoStop}%
\bibitem [{\citenamefont {Berger}\ \emph {et~al.}(1984)\citenamefont {Berger},
  \citenamefont {Girod},\ and\ \citenamefont {Gogny}}]{Berger84}%
  \BibitemOpen
  \bibfield  {author} {\bibinfo {author} {\bibfnamefont {J.}~\bibnamefont
  {Berger}}, \bibinfo {author} {\bibfnamefont {M.}~\bibnamefont {Girod}}, \
  and\ \bibinfo {author} {\bibfnamefont {D.}~\bibnamefont {Gogny}},\ }\href
  {\doibase https://doi.org/10.1016/0375-9474(84)90240-9} {\bibfield  {journal}
  {\bibinfo  {journal} {Nuclear Physics A}\ }\textbf {\bibinfo {volume}
  {428}},\ \bibinfo {pages} {23} (\bibinfo {year} {1984})}\BibitemShut
  {NoStop}%
\bibitem [{\citenamefont {Sheikh}\ \emph {et~al.}(2021)\citenamefont {Sheikh},
  \citenamefont {Dobaczewski}, \citenamefont {Ring}, \citenamefont {Robledo},\
  and\ \citenamefont {Yannouleas}}]{Sheikh21a}%
  \BibitemOpen
  \bibfield  {author} {\bibinfo {author} {\bibfnamefont {J.~A.}\ \bibnamefont
  {Sheikh}}, \bibinfo {author} {\bibfnamefont {J.}~\bibnamefont {Dobaczewski}},
  \bibinfo {author} {\bibfnamefont {P.}~\bibnamefont {Ring}}, \bibinfo {author}
  {\bibfnamefont {L.~M.}\ \bibnamefont {Robledo}}, \ and\ \bibinfo {author}
  {\bibfnamefont {C.}~\bibnamefont {Yannouleas}},\ }\href {\doibase
  10.1088/1361-6471/ac288a} {\bibfield  {journal} {\bibinfo  {journal} {Journal
  of Physics G: Nuclear and Particle Physics}\ }\textbf {\bibinfo {volume}
  {48}},\ \bibinfo {pages} {123001} (\bibinfo {year} {2021})}\BibitemShut
  {NoStop}%
\bibitem [{\citenamefont {Bally}\ and\ \citenamefont
  {Bender}(2021)}]{Bally21b}%
  \BibitemOpen
  \bibfield  {author} {\bibinfo {author} {\bibfnamefont {B.}~\bibnamefont
  {Bally}}\ and\ \bibinfo {author} {\bibfnamefont {M.}~\bibnamefont {Bender}},\
  }\href {\doibase 10.1103/PhysRevC.103.024315} {\bibfield  {journal} {\bibinfo
   {journal} {Phys. Rev. C}\ }\textbf {\bibinfo {volume} {103}},\ \bibinfo
  {pages} {024315} (\bibinfo {year} {2021})}\BibitemShut {NoStop}%
\bibitem [{\citenamefont {Anguiano}\ \emph {et~al.}(2001)\citenamefont
  {Anguiano}, \citenamefont {Egido},\ and\ \citenamefont
  {Robledo}}]{Anguiano01a}%
  \BibitemOpen
  \bibfield  {author} {\bibinfo {author} {\bibfnamefont {M.}~\bibnamefont
  {Anguiano}}, \bibinfo {author} {\bibfnamefont {J.}~\bibnamefont {Egido}}, \
  and\ \bibinfo {author} {\bibfnamefont {L.}~\bibnamefont {Robledo}},\ }\href
  {\doibase https://doi.org/10.1016/S0375-9474(01)01219-2} {\bibfield
  {journal} {\bibinfo  {journal} {Nuclear Physics A}\ }\textbf {\bibinfo
  {volume} {696}},\ \bibinfo {pages} {467 } (\bibinfo {year}
  {2001})}\BibitemShut {NoStop}%
\bibitem [{\citenamefont {Balian}\ and\ \citenamefont
  {Br{\'e}zin}(1969)}]{Balian69a}%
  \BibitemOpen
  \bibfield  {author} {\bibinfo {author} {\bibfnamefont {R.}~\bibnamefont
  {Balian}}\ and\ \bibinfo {author} {\bibfnamefont {E.}~\bibnamefont
  {Br{\'e}zin}},\ }\href@noop {} {\bibfield  {journal} {\bibinfo  {journal}
  {Nuovo Cimento}\ }\textbf {\bibinfo {volume} {64}},\ \bibinfo {pages} {37}
  (\bibinfo {year} {1969})}\BibitemShut {NoStop}%
\bibitem [{\citenamefont {Bally}\ \emph {et~al.}(2021)\citenamefont {Bally},
  \citenamefont {S{\'a}nchez-Fern{\'a}ndez},\ and\ \citenamefont
  {Rodr{\'i}guez}}]{Bally21a}%
  \BibitemOpen
  \bibfield  {author} {\bibinfo {author} {\bibfnamefont {B.}~\bibnamefont
  {Bally}}, \bibinfo {author} {\bibfnamefont {A.}~\bibnamefont
  {S{\'a}nchez-Fern{\'a}ndez}}, \ and\ \bibinfo {author} {\bibfnamefont
  {T.~R.}\ \bibnamefont {Rodr{\'i}guez}},\ }\href {\doibase
  10.1140/epja/s10050-021-00369-z} {\bibfield  {journal} {\bibinfo  {journal}
  {The European Physical Journal A}\ }\textbf {\bibinfo {volume} {57}},\
  \bibinfo {pages} {69} (\bibinfo {year} {2021})}\BibitemShut {NoStop}%
\bibitem [{\citenamefont {Caurier}\ \emph {et~al.}(2007)\citenamefont
  {Caurier}, \citenamefont {Men\'endez}, \citenamefont {Nowacki},\ and\
  \citenamefont {Poves}}]{Caurier07}%
  \BibitemOpen
  \bibfield  {author} {\bibinfo {author} {\bibfnamefont {E.}~\bibnamefont
  {Caurier}}, \bibinfo {author} {\bibfnamefont {J.}~\bibnamefont {Men\'endez}},
  \bibinfo {author} {\bibfnamefont {F.}~\bibnamefont {Nowacki}}, \ and\
  \bibinfo {author} {\bibfnamefont {A.}~\bibnamefont {Poves}},\ }\href
  {\doibase 10.1103/PhysRevC.75.054317} {\bibfield  {journal} {\bibinfo
  {journal} {Phys. Rev. C}\ }\textbf {\bibinfo {volume} {75}},\ \bibinfo
  {pages} {054317} (\bibinfo {year} {2007})}\BibitemShut {NoStop}%
\bibitem [{\citenamefont {Sienko}\ \emph {et~al.}(2003)\citenamefont {Sienko},
  \citenamefont {Lister},\ and\ \citenamefont {Kaye}}]{Sienko03}%
  \BibitemOpen
  \bibfield  {author} {\bibinfo {author} {\bibfnamefont {T.~A.}\ \bibnamefont
  {Sienko}}, \bibinfo {author} {\bibfnamefont {C.~J.}\ \bibnamefont {Lister}},
  \ and\ \bibinfo {author} {\bibfnamefont {R.~A.}\ \bibnamefont {Kaye}},\
  }\href {\doibase 10.1103/PhysRevC.67.064311} {\bibfield  {journal} {\bibinfo
  {journal} {Phys. Rev. C}\ }\textbf {\bibinfo {volume} {67}},\ \bibinfo
  {pages} {064311} (\bibinfo {year} {2003})}\BibitemShut {NoStop}%
\bibitem [{\citenamefont {Duguet}\ \emph {et~al.}(2003)\citenamefont {Duguet},
  \citenamefont {Bender}, \citenamefont {Bonche},\ and\ \citenamefont
  {Heenen}}]{Duguet03}%
  \BibitemOpen
  \bibfield  {author} {\bibinfo {author} {\bibfnamefont {T.}~\bibnamefont
  {Duguet}}, \bibinfo {author} {\bibfnamefont {M.}~\bibnamefont {Bender}},
  \bibinfo {author} {\bibfnamefont {P.}~\bibnamefont {Bonche}}, \ and\ \bibinfo
  {author} {\bibfnamefont {P.-H.}\ \bibnamefont {Heenen}},\ }\href {\doibase
  https://doi.org/10.1016/S0370-2693(03)00330-7} {\bibfield  {journal}
  {\bibinfo  {journal} {Physics Letters B}\ }\textbf {\bibinfo {volume}
  {559}},\ \bibinfo {pages} {201} (\bibinfo {year} {2003})}\BibitemShut
  {NoStop}%
\bibitem [{\citenamefont {Egido}()}]{EgidoPriv}%
  \BibitemOpen
  \bibfield  {author} {\bibinfo {author} {\bibfnamefont {J.~L.}\ \bibnamefont
  {Egido}},\ }\href@noop {} {\bibinfo  {journal} {private communication}\
  }\BibitemShut {NoStop}%
\end{thebibliography}%

\end{document}